\documentclass[11pt]{article}
\usepackage{moriond,epsfig}

\def\be{\begin{equation}}
\def\ee{\end{equation}}
\def\bea{\begin{eqnarray}}
\def\eea{\end{eqnarray}}

\begin{document}
\vspace*{4cm}
\title{Looking for non-Gaussianity in the COBE-DMR data with spherical wavelets}

\author{R.B.BARREIRO$^1$, M.P.HOBSON$^1$, A.N.LASENBY$^1$, A.J.BANDAY$^2$,
K.M. G\'ORSKI$^3$ \\ and G. HINSHAW$^4$}

\address{$^1$ Astrophysics Group, Cavendish Laboratory, Madingley Road,
Cambridge CB3 0HE, UK \\
$^2$ Max-Planck Institut fuer Astrophysik (MPA), Karl-Schwarzschild
Str.1, D-85740, Garching, Germany \\
$^3$ European Southern Observatory (ESO), Karl-Schwarzschild
Str.2, D-85740, Garching, Germany \\
$^4$ NASA/GSFC, Greenbelt, MD 20771, USA }

\maketitle\abstracts{
We present an analysis of the Gaussianity of the 4-year COBE-DMR data (in HEALPix
pixelisation) based on spherical wavelets.
The skewness, kurtosis and scale-scale correlation spectra are
computed from the detail wavelet coefficients at each scale.
The sensitivity of the method to the orientation of the data is also
taken into account. We find a single detection of non-Gaussianity at
the $>99\%$ confidence level in one of our statistics.
We use Monte-Carlo simulations to assess the statistical
significance of this detection and find that the probability of
obtaining such a detection by chance for an underlying Gaussian field
is as high as $0.69$. Therefore, our analysis does not show
evidence of non-Gaussianity in the COBE-DMR data.}

\section{Introduction}
Testing the Gaussianity of the cosmic microwave background (CMB)
fluctuations has become of great interest since it would make it possible to
distinguish between competing theories of structure formation in the
early Universe such as inflation and topological defects.
With this aim, a large number of techniques have already been
proposed, many of them being
applied to the 4-year COBE-DMR data (see e.g. Barreiro 2000). Although
most of these analyses did not find evidence for non-Gaussianity,
methods based on bispectrum analyses (Ferreira et
al. 1998, Magueijo 1999; see also Zaroubi et al. 1999) and on
statistics of wavelet coefficients
(Pando et al. 1998) have yielded detection of non-Gaussianity in the
COBE-DMR data.
In particular, Pando et al. have found a significant non-Gaussian
signal at the $99\%$ confidence level on computing the
scale-scale correlation of the wavelet coefficients. This analysis
has recently been revised by Mukherjee et al. (2000) (hereinafter MHL)
who take into account that the results depend critically on the
orientation of the signal and that a large number of the computed
statistics do not show deviation from the Gaussian case.
According to MHL, Gaussianity can be ruled out only at the $41\%$
confidence level in the DSMB data and at the $72\%$ level in the
53+90 GHz coadded data.
The above wavelet analyses have been performed applying planar
wavelets to Face 0 and Face 5 of the COBE-DMR QuadCube
pixelisation. Therefore only one-third of the available data is being
considered and, at the same time, undesirable projection effects may be
present. In this work, we have performed a similar analysis to those
of Pando et al and MHL, but applying orthogonal spherical Haar
wavelets (SHW) (see Sweldens 1995 and references therein) 
to the COBE data in HEALPix pixelisation (G\'orski et al 1999)
with the customised Galactic cut (Banday et al. 1997). 
On the one hand, this hierarchical
pixelisation scheme is particularly well-suited to the application of
such a wavelet decomposition. On the other hand, SHW
are more appropriate to study data over a large region of the
sky and also allow an easy identification of those
coefficients affected by the Galaxy. Therefore {\em all} data lying
outside the Galactic cut is used in the analysis, i.e.,
approximately two-thirds of the total number of COBE-DMR pixels.

As an illustration, we present our results for three different orientations of
the data to account for the sensitivity of these estimators to the
orientation of the input signal.

\section{The wavelet analysis}
We have used  the coadded 53+90 GHz COBE-DMR
map (each weighted according to the inverse of its noise variance)
in HEALPix pixelisation (see Banday et
al. 2000 for the map-making) with the customised Galactic cut. 
HEALPix is an equal area, iso-latitude and hierarchical
pixelisation of the sphere. The resolution
level of the grid can be expressed by a parameter $j$ (or equivalently
$N_{side}$ with $N_{side}=2^{j-1}$).
The lowest resolution ($j=1$) comprises twelve
pixels in three rings around the poles and equator. 
To move to a higher resolution level, each pixel at resolution $j$
is divided into four pixels at resolution $j+1$. Therefore
the total number of pixels at a given resolution $j$ is 
$n_j=12 \times 4^{j-1}$. The COBE-DMR maps correspond to a resolution $J=7$.

Orthogonal SHW are not translations and dilations of a
given function and so can be adapted to more general spaces
than $R^n$ (Sweldens 1995). In addition, 
they still enjoy the usual properties of planar wavelets such as a good
frequency-space localisation and a fast transform algorithm.
The temperature field can be decomposed in terms of the SHW basis
functions as
\begin{equation}
\frac{\Delta T}{T}(x_i)=\sum_{l=0}^{n_{j_0}-1}
\lambda_{j_0,l}\varphi_{j_0,l}(x_i)+
\sum_{j=j_0}^{J-1}\sum_{m=1}^3\sum_{l=0}^{n_j-1}\gamma_{m,j,l}\psi_{m,j,l}(x_i)
\end{equation} 
where $\lambda_{j_0,l}$ and $\gamma_{j,m,l}$ are the approximation and
detail wavelet coefficients respectively. 
The first term in the previous equation corresponds to a coarser
resolution image of the original map whereas the detail coefficients
encode the difference between both maps.
The index $j$ runs from the lowest resolution considered $j_0$ to $J-1$.
Finally, the index $m$ corresponds to the three different wavelet
functions at each scale needed to have a wavelet basis.
In order to get the wavelet coefficients, one starts identifying the
pixels at the map resolution $J$ with the
approximation coefficients $\lambda_{J,l}$. 
Then each wavelet coefficient at resolution $J-1$ is
obtained as a linear combination of the four corresponding coefficients 
at resolution $J$. In this way, the approximation
coefficients $\lambda_{J-1,l}$ and three sets of detail
coefficients $\gamma_{1,J-1,l}$, $\gamma_{2,J-1,l}$ and $\gamma_{3,J-1,l}$ 
at scale $J-1$ are obtained.
The same process is performed again but starting with the $\lambda_{J-1,l}$ 
coefficients as our initial map to obtain the wavelet coefficients at
$J-2$ and then the process is repeated down to the lowest resolution 
considered $j_0$ (for a more detailed description see Barreiro et al. 2000,
Tenorio et al. 1999).

In our wavelet analysis, we obtain estimators for the skewness
$\hat{S}(j,m)$, (excess) kurtosis $\hat{K}(j,m)$ and scale-scale
correlation $\hat{C}^2(j,m)$ (as defined in Barreiro et
al. 2000) of each type $m$ of detail wavelet coefficients at each
scale $j$. In order to perform our non-Gaussianity test, we first 
obtain the previous quantities for the coadded 53+90 GHz
COBE-DMR map, discarding those wavelet coefficients coming from pixels
inside the Galactic cut.
Then the same estimators are obtained for a large number (10000) of
CMB all-sky simulated maps, that take
into account the characteristics of the COBE-DMR data.
In this way, we can obtain approximate probability distributions for the 
$\hat{S}(j,m)$, $\hat{K}(j,m)$ and $\hat{C}^2(j,m)$ statistics for a 
CMB signal derived from the chosen model. In particular, the CMB 
simulations are
drawn from a inflationary/CDM model with parameters $\Omega_m=1$,
$\Omega_\Lambda=0$, $h=0.5$, $n=1$ and $Q_{rms-ps}=18\mu K$ but we do
not expect our results to vary significantly with a different
choice of parameters (see MHL).
By comparing these distributions with the values computed from the
COBE-DMR data, we can obtain the probability that our data are derived
from a Gaussian distribution characterised by the chosen power spectrum.
In addition, since the wavelet
decomposition is sensitive to the orientation of the data, we 
perform the previous test for three different orientations of the
sky. Therefore we compute a total of $42 \times 3 = 126$ statistics
for each map.

\begin{figure}[t]
\psfig{figure=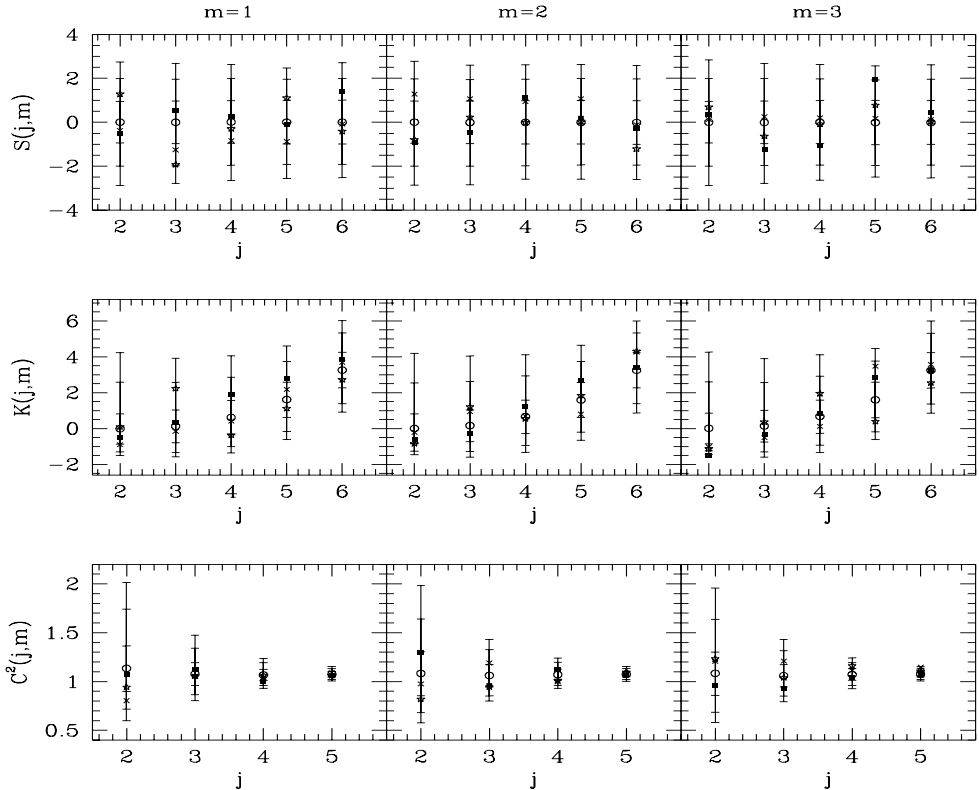
,height=11.2cm,width=14cm
}
\caption{The skewness, kurtosis and scale-scale correlation spectra
for the 53+90 GHz COBE map
are plotted for each type of detail coefficients $m=1,2,3$ (first,
second and third column respectively). 
See text for details.
\label{spectra}}
\end{figure}
\section{Results}
The computed $\hat{S}(j,m)$, $\hat{K}(j,m)$ and $\hat{C}^2(j,m)$
spectra are plotted in Fig.~\ref{spectra} for three different
orientations. We rotate the 53+90 COBE-DMR map around an axis passing
through the North and South Galactic poles. A rotation of $90$ degrees
around this direction simply shifts the twelve base-pixels of the
HEALPix pixelisation into each other and therefore gives rise to the 
same results as the original orientation.
We show the results computed from the COBE-DMR data for a
rotation of 0 (orientation A, solid squares), 30 (orientation B,
crosses) and 60 (orientation C, stars) degrees with respect to the 
original orientation of the map. The open circle and error bars correspond to
the average value and the 68, 95 and $99\%$
confidence levels of the distribution obtained from the 
10000 CDM realisations for orientation A. We do not plot the
corresponding error bars for orientations B and C in Fig.~\ref{spectra} 
for the sake of clarity and because the conclusions derived from the plot
remain unchanged.
For convenience, the skewness and kurtosis spectra
have been normalised at each value of $(j,m)$ such that
the variance of the distribution obtained from the 10000 CDM
realisations is equal to unity. 
We can see in Fig.~\ref{spectra} that the COBE-DMR values of the
$\hat{S}(j,m)$ and $\hat{C}^2(j,m)$ spectra
for orientations A,B and C 
are consistent with being derived from a parent Gaussian distribution.
However, we find a detection of non-Gaussianity in the $\hat{K}(j,m)$ 
spectra at $j=2$ and $m=3$ at the $>99\%$ 
confidence level for orientation A, the rest of the values lying
within their respective Gaussian probability distributions.

In order to assess the statistical significance of this
detection and since most of our statistics show no evidence of
non-Gaussianity, we have used Monte-Carlo simulations to estimate the
probability of having at least one detection of non-Gaussianity at the
$\ge 99\%$ confidence level out of
our 126 statistics for an underlying Gaussian field.
We find that this occurs in $69\%$ of the cases and
therefore this analysis does not provide strong evidence of
non-Gaussianity in the COBE-DMR data. This is in agreement with the results
obtained by MHL.

We have also checked that the general conclusions regarding
non-Gaussianity are not affected, at least in our case, by the
choice of a different set of orientations (see Barreiro et al. 2000).

\section{Conclusions}
We have investigated the Gaussianity of the 4-year COBE data 
(in HEALPix pixelisation) with an analysis based on SHW. 
We have taken into account the
sensitivity of our method with respect to the orientation of the input
signal, presenting the results for three different orientations.

We have found a single detection of non-Gaussianity at the $>99\%$
confidence level out of our 126 computed statistics, corresponding to
the value of the kurtosis at $j=2,m=3$ for one of the chosen
orientations. Using Monte-Carlo simulations we estimate that the
probability of having such a detection
in one of our statistics is as high as
$0.69$ for the case of an underlying Gaussian field. Therefore,
we conclude that an analysis based on SHW of the 4-year COBE-DMR data show
no evidence of non-Gaussianity in the CMB.

\section*{Acknowledgements}
RBB thanks Luis Tenorio for helpful comments about spherical wavelets.
We thank all the people involved in producing the HEALPix package, which
has been extensively used along this work.
RBB acknowledges financial support from the PPARC in the form of a
research grant.

\end{document}